\documentclass{article}
\usepackage{amsmath,amssymb}
\newcommand{\bs}[1]{\boldsymbol{#1}}

\renewcommand{\i}{\mathrm{i}}
\renewcommand{\d}{\mathrm{d}}

\newcommand{\rf}[1]{${}^{[#1]}$}
\newcommand{\rff}[1]{${}^{[#1],}$}
\begin{document}
\begin{center}
\textbf{\Large THE DUAL CANONICAL FORMALISM IN THE QUANTUM JUNCTION SYSTEMS}
\end{center}
\begin{center}
MORISHIGE YONEDA\footnote{Denshi Gakuen Japan Electronics College,
1-25-4 Hyakunin-cho, Shinjuku-Ku, Japan yoneda@jec.ac.jp},MASAAKI NIWA\footnote{Tokyo Denki University Department of Physics,2-2 Kanda-Nishiki-cho,Chiyoda-ku,Tokyo, Japan niwa@cck.dendai.ac.jp
} AND MITSUYA MOTOHASHI\footnote{Tokyo Denki University Department of Information and Communication Engineering,2-2 Kanda-Nishiki-cho, Chiyoda-ku, Tokyo, Japan mmitsuya@cck.dendai.ac.jp}
\end{center}
We have constructed a theory of dual canonical formalism, to study the quantum dual systems. In such a system, as the relationship between current and voltage of each, we assumed the duality conditions. As a simple example, we examined the duality between the SC(superconductor)/SI(superinsulator)/SC(superconductor) junction and its reverse SI/SC/SI junction. We derived the relationship between the phase and the number of particles in a dual system of each other.
\section{Introduction}
The duality has been known to be a powerful tool in various physical systems such as statistical mechanics\rf{1} and field theory\rff{2}\rf{3}. Furthermore, the duality has been also known in several studies of the Josephson junction systems\rff{4}\rf{5}.In superconducting systems, the general Josephson junction are known as the quantum effect devices\rff{6}\rf{7} which operates by using the Cooper-pair tunneling. Furthermore, the mesoscopic Josephson junction can be used as a quantum effect devices which operates by using the single Cooper-pair tunneling created by a Coulomb blockade\rf{8}. In contrast, the mesoscopic dual Josephson junction systems are known as the quantum effect devices which operates by using the single quantum vortex(quantum flux) tunneling\rf{9}. The Cooper-pair and the quantum vortex are known as the duality relation to each other\rff{10}\rf{11}. That is, under certain conditions, even if we replace the role of Cooper pairs and vortices, these systems satisfy the equation of similar type. Therefore, these two devices can be understood as a dual device with each other. However, a clear definition between such a dual system  has not yet been established in general. So we are to study the dual quantum system, we have constructed a theory of dual canonical formalism. This paper is composed as follows. In the next section, as the simplest example, we show the dual canonical formalism of harmonic oscillator. In Sec.3, we build the dual canonical form of the quantum LC circuit. In Sec.4, we formulate the theoretical model and basic equations for single SC(supercnductor)/SI(superinsulator)/SC(supercnductor) junction and single SI/SC/SI junction. In addition, we derived the quantum resistance by the two ways method. In Sec.5, as a generalization of the previous chapters, we introduce the superconductor-superinsulator network system of compete with each other and then summarize results in Sec.6.
\section{Dual canonical formalism of the harmonic oscillator}
In this section, as the simplest example, we show the dual canonical formalism of harmonic oscillator. First, we introduce the following commutation relations of between the coordinate$\,x(t)$ and momenta$\,p(t)$
\begin{equation}
\Bigl[\,x(t),\ p(t^{\prime})\,\Bigr]_-\equiv x(t)p(t^{\prime})- p(t^{\prime})x(t)=\i\hbar\delta(t-t^{\prime}),\label{2.1}
\end{equation}
The Hamiltonian$\,H$ of  the harmonic oscillator is given by
\begin{equation}
H=\frac{1}{2m}p^2(t)+\frac{m\omega^2}{2}x^2(t),\label{2.2}
\end{equation}
where $m$ and $\omega$ are mass and angular frequency of harmonic oscillator respectively. From the Heisenberg's equation of motion, the equations of motion are given by
\begin{subequations}
\begin{align}
\dot{x}&\equiv\frac{\d x(t)}{\d t}=\frac{\i}{\hbar}\Bigl[\,H,\ x(t)\,\Bigr]_-=\frac{p(t)}{m},\label{2.3a}\\
f(t)&\equiv\frac{\d p(t)}{\d t}=\frac{\i}{\hbar}\Bigl[\,H,\ p(t)\,\Bigr]_-=-m\omega^2x(t),\label{2.3b}
\end{align}
\end{subequations}
where $\dot{x}(t)$ and $f(t)$ are velocity and force of harmonic oscillator respectively. 

As a next step, we introduce the following commutation relations of between the dual coordinate $\widetilde{x}(t)$ and dual momenta $\widetilde{p}(t)$ 	
 \begin{equation}
\Bigl[\,\widetilde{x}(t),\ \widetilde{p}(t^{\prime})\,\Bigr]_-=\i\hbar\delta(t-t^{\prime}).\label{2.4}
\end{equation}
Then the dual Hamiltonian $\widetilde{H}$ of  eq.(2) is given by 
\begin{equation}
\widetilde{H}=\frac{1}{2\widetilde{m}}\widetilde{p}^2(t)+\frac{\widetilde{m}\widetilde{\omega}^2}{2}\widetilde{x}^2(t),\label{2.5}
\end{equation}
where $\widetilde{m}$ and $\widetilde{\omega}$ are dual mass and dual angular frequency of harmonic oscillator respectively.

The Heisenberg's equation of motion  are given by
\begin{subequations}
\begin{align}
\dot{\widetilde{x}}&\equiv\frac{\d \widetilde{x}(t)}{\d t}=\frac{\i}{\hbar}\Bigl[\,\widetilde{H},\ \widetilde{x}(t)\,\Bigr]_-=\frac{\widetilde{p}(t)}{\widetilde{m}},\label{2.6a}\\
\widetilde{f}(t)&\equiv -\frac{\d \widetilde{p}(t)}{\d t}=-\frac{\i}{\hbar}\Bigl[\,\widetilde{H},\ \widetilde{p}(t)\,\Bigr]_-=\widetilde{m}\widetilde{\omega}^2\widetilde{x}(t).\label{2.6b}
\end{align}
\end{subequations}
where $\dot{\widetilde{x}}(t)$ and $\widetilde{f}(t)$ are dual velocity and dual force of harmonic oscillator respectively.

Between eqs.(\ref{2.3a}), (\ref{2.3b}) and eqs.(\ref{2.6a}), (\ref{2.6b}) we assume the following the duality conditions:
\begin{subequations}
\begin{align}
\dot{x}(t)&=\widetilde{f}(t),\label{2.7a}\\
f(t)&=\dot{\widetilde{x}}(t).\label{2.7b}
\end{align}
\end{subequations}
By imposing the duality conditions, we obtain the next two type relationship, One of them are relationship between coordinate and momenta 
\begin{subequations}
\begin{align}
x(t)&=-\widetilde{p}(t),\label{2.8a}\\
\widetilde{x}(t)&=p(t).\label{2.8b}
\end{align}
\end{subequations}
Another one of them are relationship between mass and angular frequency  
\begin{subequations}
\begin{align}
\widetilde{m}&=\frac{1}{m\omega^2},\label{2.9a}\\
\frac{1}{m}&=\widetilde{m}\widetilde{\omega}^2\label{2.9b}
\end{align}
\end{subequations}
From the above, we derived the duality conditions and relationship of relations of variables and constants. In particular, the duality conditions of (\ref{2.7a}) and (\ref{2.7b}) are important. These conditions can also be applied to the harmonic oscillators as well as general dynamical systems.
\section{Dual canonical formalism of the quantum LC circuit}
In this section, using the methods of the previous section, we build the dual canonical form of the quantum LC circuit, the commutation relations of between the electric charge $Q(x)$ and magnetic flux $\varPhi(x)$ are as follows:
\begin{equation}
\Bigl[\,\varPhi(x),\,Q(x^{\prime})\,\Bigr]_-=\i\hbar\delta(x-x^{\prime}),\label{3.1}
\end{equation}
The Hamiltonian $H_{\mathrm{LC}}$ of the quantum LC circuit is given by
\begin{equation}
H_{\mathrm{LC}}=\frac{1}{2C}Q^2(t)+\frac{1}{2L}\varPhi^2(t),\label{3.2}
\end{equation}
where $C$ and $L$ are capacitance and inductance of LC circuit respectively. The Heisenberg's equations of motion are given by
\begin{subequations}
\begin{align}
I(t)&\equiv\frac{\d Q(t)}{\d t}=\frac{\i}{\hbar}\Bigl[\,H_{\mathrm{LC}},\ Q(t)\,\Bigr]_-=-\frac{\varPhi(t)}{L},\label{3.3a}\\
V(t)&\equiv -\frac{\d \varPhi(t)}{\d t}=\frac{\i}{\hbar}\Bigl[\,H_{\mathrm{LC}},\ \varPhi(t)\,\Bigr]_-=\frac{Q(t)}{C}.\label{3.3b}
\end{align}
\end{subequations}
where $I(t)$ and $V(t)$ are current and voltage of LC circuit respectively. 

As a next step, we introduce a dual LC circuit system and construct a commutation relation of between the dual electric charge $\widetilde{Q}(x)$ and dual magnetic flux $\widetilde{\varPhi}(x)$ as 
\begin{equation}
\Bigl[\,\widetilde{\varPhi}(x),\,\widetilde{Q}(x^{\prime})\,\Bigr]_-=\i\hbar\delta(x-x^{\prime}).\label{3.4}
\end{equation}
The dual Hamiltonian $\widetilde{H}_{\mathrm{LC}}$ of  eq.(\ref{3.2})  is given by
\begin{equation}
\widetilde{H}_{\mathrm{LC}}=\frac{1}{2\widetilde{C}}\widetilde{Q}^2(t)+\frac{1}{2\widetilde{L}}\widetilde{\varPhi}^2(t),\label{3.5}
\end{equation}
where $\widetilde{C}$ and $\widetilde{L}$ are a dual capacitance and  a dual inductance of the dual LC circuit respectively. The Heisenberg's equations of motion are given by
\begin{subequations}
\begin{align}
\widetilde{I}(t)&\equiv\frac{\d \widetilde{Q}(t)}{\d t}=\frac{\i}{\hbar}\Bigl[\,\widetilde{H}_{\mathrm{LC}},\ \widetilde{Q}(t)\,\Bigr]_-=-\frac{\widetilde{\varPhi}(t)}{\widetilde{L}},\label{3.6a}\\
\widetilde{V}(t)&\equiv \frac{\d \widetilde{\varPhi}(t)}{\d t}=\frac{\i}{\hbar}\Bigl[\,\widetilde{H}_{\mathrm{LC}},\ \widetilde{\varPhi}(t)\,\Bigr]_-=\frac{\widetilde{Q}(t)}{\widetilde{C}}.\label{3.6b}
\end{align}
\end{subequations}
where $\widetilde{I}(t)$ and $\widetilde{V}(t)$ are dual current and dual voltage of dual LC circuit respectively. Between eqs.(\ref{3.3a}), (\ref{3.3b}) and eqs.(\ref{3.6a}),(\ref{3.6b}), we assume the following duality conditions:
\begin{subequations}
\begin{align}
\widetilde{I}(t)&=V(t),\label{3.7a}\\
\widetilde{V}(t)&=I(t).\label{3.7b}
\end{align}
\end{subequations}
By imposing the duality conditions, we derived the next two type relationship, One of them are relationship between charge and  flux 
\begin{subequations}
\begin{align}
\varPhi(t)&=-\widetilde{Q}(t),\label{3.8a}\\
\widetilde{\varPhi}(t)&=Q(t)\label{3.8b}
\end{align}
\end{subequations}
Another one of them are relationship between capacitance and inductance
\begin{subequations}
\begin{align}
\widetilde{C}&=L,\label{3.9a}\\
\widetilde{L}&=-C\label{3.9b}
\end{align}
\end{subequations} 
The duality conditions of (\ref{3.7a}) and (\ref{3.7b}) are is very important, when we will treat with the Josephson junction systems in the subsequent sections.
\clearpage\noindent
\section{Dual canonical formalism of between the single SC/SI/SC junction and the single SI/SC/SI junction}

Based on the results of the previous sections, we examined the duality\rff{10}\rff{12}\rf{13} between the SC/SI/SC junction and SI/SC/SI junction. First, we propose a small SC/SI/SC junction\rf{9} and its equivalent circuit are shown in Fig.1. 

\begin{figure}[h]
\unitlength 0.1in
\begin{picture}(50.00,12.25)(2.00,-14.25)
%
\special{pn 8}%
\special{pa 400 200}%
\special{pa 1600 200}%
\special{pa 1600 800}%
\special{pa 400 800}%
\special{pa 400 200}%
\special{fp}%
%
\special{pn 8}%
\special{sh 0.300}%
\special{pa 1600 800}%
\special{pa 2000 800}%
\special{pa 2000 200}%
\special{pa 1600 200}%
\special{pa 1600 800}%
\special{fp}%
%
\special{pn 8}%
\special{pa 2000 200}%
\special{pa 3200 200}%
\special{pa 3200 800}%
\special{pa 2000 800}%
\special{pa 2000 200}%
\special{fp}%
\put(10.1000,-4.0000){\makebox(0,0){SC}}%
\put(26.1000,-4.0000){\makebox(0,0){SC}}%
\put(18.0000,-4.0000){\makebox(0,0){SI}}%
%
\special{pn 8}%
\special{pa 400 500}%
\special{pa 200 500}%
\special{fp}%
\special{pa 200 500}%
\special{pa 200 1200}%
\special{fp}%
\special{pa 200 1200}%
\special{pa 1760 1200}%
\special{fp}%
%
\special{pn 8}%
\special{pa 3200 500}%
\special{pa 3400 500}%
\special{fp}%
\special{pa 3400 500}%
\special{pa 3400 1200}%
\special{fp}%
\special{pa 3400 1200}%
\special{pa 1850 1200}%
\special{fp}%
%
\special{pn 8}%
\special{pa 1750 1040}%
\special{pa 1750 1390}%
\special{fp}%
%
\special{pn 8}%
\special{pa 1860 1120}%
\special{pa 1860 1290}%
\special{fp}%
\put(10.0000,-6.0000){\makebox(0,0){$\theta_1$}}%
\put(26.0000,-6.0000){\makebox(0,0){$\theta_2$}}%
\put(17.9000,-6.0000){\makebox(0,0){$C_{\mathrm{J}}$}}%
\put(18.0000,-15.1000){\makebox(0,0){$V_{\mathrm{g}}$}}%
\put(38.3000,-7.7000){\makebox(0,0){\Large{$=$}}}%
%
\special{pn 8}%
\special{pa 4400 635}%
\special{pa 5030 635}%
\special{pa 5030 425}%
\special{pa 4400 425}%
\special{pa 4400 635}%
\special{fp}%
%
\special{pn 8}%
\special{sh 0.300}%
\special{pa 4660 425}%
\special{pa 4730 425}%
\special{pa 4730 635}%
\special{pa 4660 635}%
\special{pa 4660 425}%
\special{fp}%
%
\special{pn 8}%
\special{pa 4400 535}%
\special{pa 4200 535}%
\special{fp}%
\special{pa 4200 535}%
\special{pa 4200 965}%
\special{fp}%
\special{pa 4200 965}%
\special{pa 4680 965}%
\special{fp}%
%
\special{pn 8}%
\special{pa 5030 545}%
\special{pa 5200 545}%
\special{fp}%
\special{pa 5200 545}%
\special{pa 5200 965}%
\special{fp}%
\special{pa 5200 965}%
\special{pa 4730 965}%
\special{fp}%
%
\special{pn 8}%
\special{pa 4680 865}%
\special{pa 4680 1085}%
\special{fp}%
%
\special{pn 8}%
\special{pa 4740 915}%
\special{pa 4740 1025}%
\special{fp}%
\put(47.1000,-11.6500){\makebox(0,0){$V_{\mathrm{g}}$}}%
\put(46.9000,-3.2500){\makebox(0,0){$C_{\mathrm{J}}$}}%
\end{picture}%
\caption{Schematic of superconductor(SC)/superinsulator(SI)/superconductor(SC) junction and its equivalent circuit}
\end{figure}
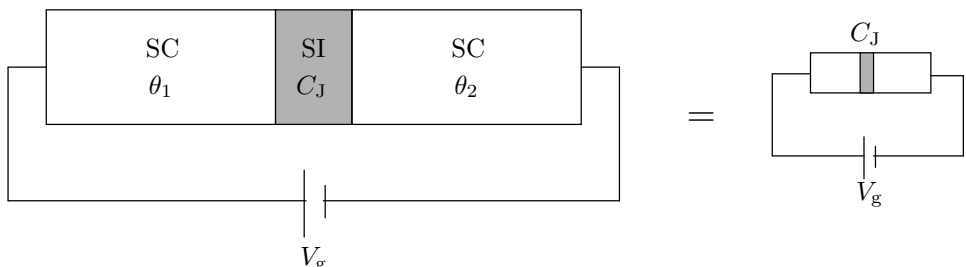

Now we introduce an order field (original particles field) $\varPsi(x)$
\begin{equation}
\varPsi(x)=\sqrt{N(x)}\exp\left(\i\theta(x)\right).\label{4.1}
\end{equation}

For example, in superconducting systems, $N=N_1-N_2$ is represent the relative number operator of Cooper pair, and $\theta=\theta_1-\theta_2$ is represent the relative phase of Cooper pair. The commutation relations between $N(x)$ and $\theta(x^{\prime})$ are given by
\begin{equation}
\Bigl[\,N(x),\,\theta(x^{\prime})\,\Bigr]_-=\i\hbar\delta(x-x^{\prime}).\label{4.2}
\end{equation}
The junction is characterized by its capacitance $C_{\mathrm{J}}$ and the Josephson energy $E_{\mathrm{J}}=\hbar I_{\mathrm{C}}/2e$, where $I_{\mathrm{C}}$ is critical current. Now, this junctions system when considered as the Josephson junctions, This system can be described by the following Hamiltonian of Cooper pair:
\begin{equation}
H_{\mathrm{J}}=4N^2E_{\mathrm{C}}+E_{\mathrm{J}}\left(1-\cos\theta\right).\label{4.3}
\end{equation}
The first term describe the Coulomb energy of Josephson junction, $N$ is the number operator of Cooper pair, and $E_{\mathrm{C}}=e^2/(2C_{\mathrm{J}})$ is charging energy per single-charge . The second term in eq.(\ref{4.3}) describe the Josephson coupling energy. From eq.(\ref{4.3}) two Josephson's equations are given by
\begin{subequations}
\begin{align}
V&=\frac{\hbar}{2e}\frac{\partial\theta(t)}{\partial t}=\frac{\hbar}{2e}\frac{\i}{\hbar}\Bigl[\,H_{\mathrm{J}},\,\theta\,\Bigr]_-=\frac{4N}{e}E_{\mathrm{C}},\label{4.4a}\\
I&=2e\frac{\partial N(t)}{\partial t}=2e\frac{\i}{\hbar}\Bigl[\,H_{\mathrm{J}},\,N\,\Bigr]_-=-I_{\mathrm{C}}\sin\theta,\label{4.4b}
\end{align}
\end{subequations}
where $V$ and $I$ are voltage and current of Cooper pair, respectively. 

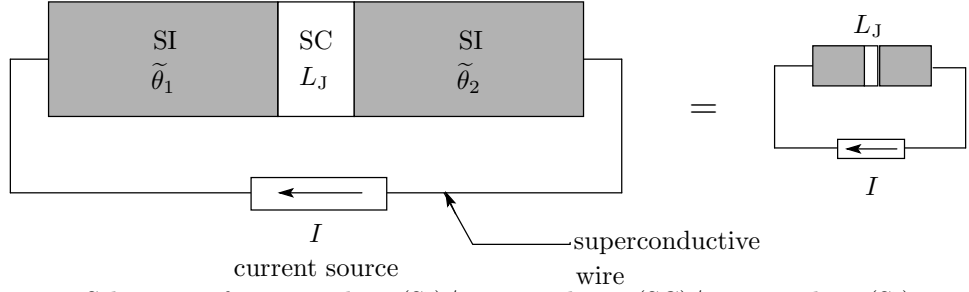
\begin{figure}
\unitlength 0.1in
\begin{picture}(51.90,13.10)(0.10,-15.10)
%
\special{pn 8}%
\special{sh 0.300}%
\special{pa 400 200}%
\special{pa 1600 200}%
\special{pa 1600 800}%
\special{pa 400 800}%
\special{pa 400 200}%
\special{fp}%
%
\special{pn 8}%
\special{pa 1600 800}%
\special{pa 2000 800}%
\special{pa 2000 200}%
\special{pa 1600 200}%
\special{pa 1600 800}%
\special{fp}%
%
\special{pn 8}%
\special{sh 0.300}%
\special{pa 2000 200}%
\special{pa 3200 200}%
\special{pa 3200 800}%
\special{pa 2000 800}%
\special{pa 2000 200}%
\special{fp}%
\put(10.1000,-4.0000){\makebox(0,0){SI}}%
\put(26.1000,-4.0000){\makebox(0,0){SI}}%
\put(18.0000,-4.0000){\makebox(0,0){SC}}%
%
\special{pn 8}%
\special{pa 400 500}%
\special{pa 200 500}%
\special{fp}%
\special{pa 200 500}%
\special{pa 200 1200}%
\special{fp}%
\special{pa 200 1200}%
\special{pa 1760 1200}%
\special{fp}%
%
\special{pn 8}%
\special{pa 3200 500}%
\special{pa 3400 500}%
\special{fp}%
\special{pa 3400 500}%
\special{pa 3400 1200}%
\special{fp}%
\special{pa 3400 1200}%
\special{pa 1850 1200}%
\special{fp}%
\put(10.0000,-6.0000){\makebox(0,0){$\widetilde{\theta}_1$}}%
\put(26.0000,-6.0000){\makebox(0,0){$\widetilde{\theta}_2$}}%
\put(17.9000,-6.0000){\makebox(0,0){$L_{\mathrm{J}}$}}%
\put(18.0000,-14.1000){\makebox(0,0){$I$}}%
\put(38.3000,-7.7000){\makebox(0,0){\Large{$=$}}}%
%
\special{pn 8}%
\special{pa 4670 430}%
\special{pa 4740 430}%
\special{pa 4740 640}%
\special{pa 4670 640}%
\special{pa 4670 430}%
\special{fp}%
%
\special{pn 8}%
\special{pa 4400 535}%
\special{pa 4200 535}%
\special{fp}%
\special{pa 4200 535}%
\special{pa 4200 965}%
\special{fp}%
\special{pa 4200 965}%
\special{pa 4680 965}%
\special{fp}%
%
\special{pn 8}%
\special{pa 5030 545}%
\special{pa 5200 545}%
\special{fp}%
\special{pa 5200 545}%
\special{pa 5200 965}%
\special{fp}%
\special{pa 5200 965}%
\special{pa 4730 965}%
\special{fp}%
\put(47.1000,-11.6500){\makebox(0,0){$I$}}%
\put(46.9000,-3.2500){\makebox(0,0){$L_{\mathrm{J}}$}}%
%
\special{pn 8}%
\special{sh 0.300}%
\special{pa 4400 430}%
\special{pa 4670 430}%
\special{pa 4670 640}%
\special{pa 4400 640}%
\special{pa 4400 430}%
\special{fp}%
%
\special{pn 8}%
\special{sh 0.300}%
\special{pa 4750 430}%
\special{pa 5020 430}%
\special{pa 5020 640}%
\special{pa 4750 640}%
\special{pa 4750 430}%
\special{fp}%
%
\special{pn 8}%
\special{sh 0}%
\special{pa 1460 1120}%
\special{pa 2170 1120}%
\special{pa 2170 1290}%
\special{pa 1460 1290}%
\special{pa 1460 1120}%
\special{fp}%
%
\special{pn 8}%
\special{pa 2040 1200}%
\special{pa 1610 1200}%
\special{fp}%
\special{sh 1}%
\special{pa 1610 1200}%
\special{pa 1677 1220}%
\special{pa 1663 1200}%
\special{pa 1677 1180}%
\special{pa 1610 1200}%
\special{fp}%
\put(18.0000,-15.9000){\makebox(0,0){current source}}%
%
\special{pn 8}%
\special{sh 0}%
\special{pa 4530 920}%
\special{pa 4880 920}%
\special{pa 4880 1020}%
\special{pa 4530 1020}%
\special{pa 4530 920}%
\special{fp}%
%
\special{pn 8}%
\special{pa 4840 970}%
\special{pa 4580 970}%
\special{fp}%
\special{sh 1}%
\special{pa 4580 970}%
\special{pa 4647 990}%
\special{pa 4633 970}%
\special{pa 4647 950}%
\special{pa 4580 970}%
\special{fp}%
%
\special{pn 8}%
\special{pa 2470 1200}%
\special{pa 2620 1470}%
\special{pa 3130 1470}%
\special{pa 3130 1460}%
\special{pa 3130 1460}%
\special{fp}%
%
\special{pn 8}%
\special{pa 2530 1310}%
\special{pa 2480 1200}%
\special{fp}%
\special{sh 1}%
\special{pa 2480 1200}%
\special{pa 2489 1269}%
\special{pa 2502 1249}%
\special{pa 2526 1252}%
\special{pa 2480 1200}%
\special{fp}%
\put(31.5000,-15.2000){\makebox(0,0)[lb]{superconductive}}%
\put(31.6000,-16.8000){\makebox(0,0)[lb]{wire}}%
\end{picture}%
\caption{Schematic of superinsulator(SI)/superconductor(SC)/superinsulator(SI) junction and its equivalent circuit}
\end{figure}
Then, we describe the theoretical model and basic equations for the single superinsulator(SI)/superconductor(SC)/superinsulator(SI) junction. A small SI/SC/SI junction and its equivalent circuit are shown in Fig.2. On the analogy of an order field(\ref{4.1}), we introduce a disorder field\rff{10}\rff{12}\rff{13}\rf{14}(dual particles field)  
\begin{equation}
\widetilde{\varPsi}(x)=\sqrt{\widetilde{N}(x)}\exp\left(\i\widetilde{\theta}(x)\right)\label{4.5}
\end{equation}
For example, in superconducting systems, $\widetilde{N}=\widetilde{N}_1-\widetilde{N}_2$ is represent the relative number operator of vortex, and $\widetilde{\theta}=\widetilde{\theta}_1-\widetilde{\theta}_2$ is represent the relative phase of vortex. We suppose that the commutation relations between $\widetilde{N}(x)$ and $\widetilde{\theta}(x^{\prime})$ are given by
\begin{equation}
\Bigl[\,\widetilde{N}(x),\,\widetilde{\theta}(x^{\prime})\,\Bigr]_-=\i\delta(x-x^{\prime}),\label{4.6}
\end{equation}
and the dual Hamiltonian\rff{15}\rf{16} of eq.(\ref{2.1}) is given by
\begin{equation}
\widetilde{H}_{\mathrm{J}}=\widetilde{N}^2E_{\mathrm{v}}+\frac{2E_{\mathrm{C}}}{\pi^2}\left(1-\cos\widetilde{\theta}\right).\label{4.7}
\end{equation}
The first term of eq.(\ref{4.7}) describes the magnetic energy of dual Josephson junction, $E_{\mathrm{v}}=\varPhi^2_0/(2L_{\mathrm{c}})$ is a magnetic energy per single-vortex, and $L_{\mathrm{c}}=\varPhi_0/(2\pi I_{\mathrm{c}})$ is a critical inductance. The second term in eq.(\ref{4.7}) describes the dual Josephson coupling energy, where $\widetilde{\theta}=\widetilde{\theta}_1-\widetilde{\theta}_2$ is dual phase difference across the junction. From eq.(\ref{4.7}) two dual Josephson's equations are given by
\begin{subequations}
\begin{align}
\widetilde{V}&=\frac{\hbar}{\varPhi_0}\frac{\partial\widetilde{\theta}}{\partial t}=\frac{\hbar}{\varPhi_0}\frac{\i}{\hbar}\Bigl[\,\widetilde{H}_{\mathrm{J}},\,\widetilde{\theta}\,\Bigr]_-=I_{\mathrm{c}}2\pi\widetilde{N},\label{4.8a}\\
\widetilde{I}&=-\varPhi_0\frac{\partial \widetilde{N}}{\partial t}=\varPhi_0\frac{\i}{\hbar}\Bigl[\,\widetilde{H}_{\mathrm{J}},\,\widetilde{N}\,\Bigr]_-=\frac{2E_{\mathrm{C}}}{\pi e}\sin \widetilde{\theta},\label{4.8b}
\end{align}
\end{subequations}
where $\widetilde{V}$ and $\widetilde{I}$  are voltage of vortex and current of vortex respectively. From the conditions of (\ref{3.7a}) and (\ref{3.7b}), we derived the next two type relationship. One of them is relationship between the phase of Cooper pair $\theta(x)$ and vortex number $\widetilde{N}(x)$ as follows:
\begin{subequations}
\begin{equation}
\widetilde{N}(x)=-\frac{1}{2\pi}\sin\theta(x).\label{4.9a}
\end{equation}
Another one of them is relationship between the phase of vortex field $\widetilde{\theta}(x)$ and Cooper pair number $N(x)$ as follows:
\begin{equation}
N(x)=\frac{1}{2\pi}\sin\widetilde{\theta}(x).\label{4.9b}
\end{equation}
\end{subequations}

From the duality conditions of (\ref{3.7a}) and (\ref{3.7b}), we derived the quantum resistance\rff{17}\rf{18} by the next two ways method. One of those ways is to use the ratio between the fluctuation of the number of Cooper pair and the fluctuation of the number of vortex, 
\begin{subequations}
\begin{align}
R& \equiv \frac{V}{I}=-\frac{\hbar }{(2e)^{2}}\frac{8N}{\sin \theta }\frac{%
E_{\mathrm{C}}}{E_{\mathrm{J}}}  \notag \\
& =\frac{\widetilde{I}}{I}=-\frac{h}{(2e)^{2}}\frac{\Delta \widetilde{N}}{%
\Delta N}  \notag \\
& =\frac{h}{(2e)^{2}}\frac{\sin \Delta \theta }{\sin \Delta \widetilde{%
\theta }}  \label{4.10a}
\end{align}
Another way of these is to use the ratio between the fluctuation of the phase of Cooper pair and the fluctuation of the phase of vortex,
we derived the quantum conductance ( dual quantum resistance $\widetilde{R}$) as follows
\begin{align}
\widetilde{R}& \equiv \frac{\widetilde{V}}{\widetilde{I}}=\frac{(2e)^{2}}{%
\hbar }\frac{\pi ^{2}\widetilde{N}}{2\sin \widetilde{\theta }}\frac{E_{%
\mathrm{J}}}{E_{\mathrm{C}}}  \notag \\
& =\frac{\widetilde{V}}{V}=\frac{(2e)^{2}}{\hbar }\frac{\Delta \widetilde{%
\theta }}{\Delta \theta }  \notag \\
& =-\frac{(2e)^{2}}{\hbar }\frac{\Delta N}{\Delta \widetilde{N}} \label{4.10b}
\end{align}
\end{subequations}
In the eqs.(\ref{4.10a}) or (\ref{4.10b}), a case of condition of $\Delta 
\widetilde{N}>\Delta N$ (ie:$\Delta \theta >\Delta \widetilde{\theta }$) or $%
E_{\mathrm{C}}>E_{\mathrm{J}}$, it becomes insulator state. In particular,
in this extreme case $R\rightarrow \infty $, it becomes superinsulator state
. If the reverse case, in the condition of $\Delta \widetilde{N}<\Delta N$
(ie:$\Delta \theta <\Delta \widetilde{\theta }$) or $E_{\mathrm{J}}>E_{%
\mathrm{C}}$ , it becomes conducting state. In particular, in this extreme
case $R\rightarrow 0$, it becomes superconductor state . As a special case
of these conditions, in the case of $\Delta \widetilde{N}\simeq \Delta N$ or 
$\Delta \widetilde{\theta }\simeq \Delta \theta $, it becomes self dual
state. In this case, the resistance $R$ becomes 
\begin{equation}
R\to R_{\mathrm{Q}}=\frac{h}{(2e)^2}\label{quantum resistance}
\end{equation}
where $R_{\mathrm{Q}}$ is a quantum resistance.
\clearpage\noindent
\section{The superconductor-superinsulator network systems of compete with each other} 
Until the previous chapter, we had been dealt with models of single junctions in order to simplify the problem. In this chapter, we had extended the theory from single junctions to periodic quantum dot network systems\rff{19}\rf{20}. Such a system, we considered two type competing systems. Two of those systems are superconducting dots systems (as shown in FIG.3(a))in superinsulator  host. The remaining two systems are superinsulating dots systems(as shown in FIG.3(b)) in superconducting host.

\begin{figure}[h]
\hspace*{-2.0cm}\input dualfig3.tex
\begin{center}
\caption{Schematic of superconductor-superinsulator network systems}{(a)\, Superconducting dots in superinsulator host.
\\
(b)\, Superinsulating dots in superconducting host.}
\end{center}
\end{figure}

The Hamiltonian (space $2d+$imaginary time) of Josephson network systems (lattice systems) by superinsulator dots in superconducting host(as FIG.3(b)) is written as
\begin{subequations}
\begin{align}
H&=4E_{\mathrm{C}}\sum_{i,j}\left(N(i)-\frac{Q(i)}{2e}\right)\left(M^{-1}\right)_{ij}\left(N(i)-\frac{Q(i)}{2e}\right)\notag\\
&+E_{\mathrm{J}}\sum_{i,j}\left\{1-\cos\left(\theta(i)-\theta(j)+\frac{2\pi}{\varPhi_0}A_{ij}\right)\right\}\label{5.1a}
\end{align}
where $M^{-1}$ is a dimensionless potential matrix, and $A_{ij}$ is a vector potential of lattice version
\begin{equation}
A_{ij}\equiv\int_i^{j}\d\bs{l}\cdot\bs{A}=-\left(\varPhi(i)-\varPhi(j)\right).\label{5.1b}
\end{equation}
\end{subequations}
\begin{subequations}
\begin{align}
V(i)&=\frac{2E_{\mathrm{C}}}{\pi e}\sum_a\left(2\pi N_{\mathrm{C}}(a)-2\pi\frac{q(a)}{2e}\right)\left(M^{-1}\right)_{aj},\label{5.2a}\\
I(i)&=-E_{\mathrm{J}}\frac{2e}{\hbar}\sum_a\sin\left(\theta(i)-\theta(a)+\frac{2\pi}{\varPhi_0}A_{ia}\right).\label{5.2b}
\end{align}
\end{subequations}
On the other hand, The Hamiltonian of dual Josephson network systems by superconducting dots in superinsulating host (as FIG.3(a)) is written as
\begin{subequations}
\begin{align}
\widetilde{H}&=2\pi^2E_{\mathrm{J}}\sum_{i,j}\left(\widetilde{N}(i)+\frac{\varPhi(i)}{\varPhi_0}\right)\left(\widetilde{M}^{-1}\right)_{ij}\left(\widetilde{N}(i)+\frac{\varPhi(i)}{\varPhi_0}\right)\notag\\
&+\frac{2}{\pi^2}E_{\mathrm{C}}\sum_{<i,j>}\left\{1-\cos\left(\widetilde{\theta}(i)-\widetilde{\theta}(j)+\frac{2\pi}{2e}\widetilde{A}_{ij}\right)\right\}\label{5.3a}
\end{align}
where $\widetilde{M}^{-1}$ is a dimensionless potential matrix, and $\widetilde{A}_{ij}$ is a dual vector potential of lattice version
\begin{equation}
\widetilde{A}_{ij}\equiv\int_i^{j}\d\bs{l}\cdot\widetilde{\bs{A}}=-\left(\widetilde{\varPhi}(i)-\widetilde{\varPhi}(j)\right)= -\left(Q(i)-Q(j)\right).\label{5.3b}
\end{equation}
\end{subequations}
The equations of motion are given by 
\begin{subequations}
\begin{align}
\widetilde{V}(i)&=E_{\mathrm{J}}\frac{2e}{\hbar}\sum_a\left(2\pi\widetilde{N}(a)+2\pi\frac{\varPhi(a)}{\varPhi_0}\right)\left(\widetilde{M}^{-1}\right)_{ai},\label{5.4a}\\
\widetilde{I}(i)&= \frac{4 E_{\mathrm{J}}}{2\pi e}\sum_a\sin\left(\widetilde{\theta}(i)-\widetilde{\theta}(a)+\frac{2\pi}{2e}\widetilde{A}_{ia}\right).\label{5.4b}
\end{align}
\end{subequations}
As with the previous chapter, from the duality conditions of (\ref{3.7a}) and (\ref{3.7b}), we derived the next two type relationship, 
\begin{subequations}
\begin{align}
N(a)&=\frac{1}{2\pi}\sum_b\left(M\right)_{ab}\sin\left(\widetilde{\theta}(b)-\widetilde{\theta}(a)+\frac{2\pi}{2e}\widetilde{A}_{ba}\right)+\frac{q(a)}{2e},\label{5.5a}\\
\widetilde{N}(a)&=-\frac{1}{2\pi}\sum_b\left(M\right)_{ab}\sin\left(\theta(b)-\theta(a)+\frac{2\pi}{\varPhi_0}A_{ba}\right)-\frac{\varPhi(a)}{\varPhi_0}.\label{5.5b}
\end{align}
\end{subequations}
The relation of (\ref{5.5a}) and (\ref{5.5b}) can be considered as a generalization of the relation of (\ref{4.9a}) and (\ref{4.9b}).
\section{Summary and Conclusion} 
In this paper, our results are described as follows. In the first result, we examined the harmonic oscillator system as the simplest dual canonical systems. Here, we imposed the dual condition the between force and velocity in a dual system in each other, and we derived two type relationship. One of them is relationship between coordinate and  momenta, another one of them is relationship between mass and angular frequency. In the second result, we studied the quantum LC circuit system as the simplest dual electromagnetic systems. Where, we imposed the dual condition the between electric current and voltage in a dual system in each other, and we derived two type relationship. One of them is relationship between charge and flux, another one of them is relationship between capacitance and inductance. In the third results, we studied the duality between the SC/SI/SC junction and SI/SC/SI junction. We imposed a dual condition similar to the second results, and we derived two type relationships. One of them is relationship between the phase of Cooper pair $\theta(x)$ and vortex number $\widetilde{N}(x)$. Another one of them is relationship between the phase of vortex field $\widetilde{\theta}(x)$ and Cooper pair number $N(x)$. In the fourth result, as a generalization of the third results, we had extended the theory from single junctions to periodic quantum dot network systems. We imposed a dual condition similar to the third results, and we derived the generalized relation of the third result. As mentioned above, our duality conditions will be become a universal and powerful tool for studying the generalized dual systems.
\section*{Reference}
\begin{enumerate}
\item\ H.A.Kramers and G.H.Wannier,\textit{Phys.Rev.},\textbf{60},252(1941)
\item\ R.Savit,\textit{Rev.Mod.Phys.},\textbf{52},453(1980)
\item\ E.Witten,\textit{Phys.Today},\textbf{50},28(1997)
\item\ Y.M.Blanter,R.Fazio and G.Sch\"{o}n,\textit{Nuc.Phy.B Pro.Suppl.},\textbf{V.58},79(1997)
\item\ R.Fazio and G.Sch\"{o}n,\textit{Phys.Rev.B},\textbf{43},5307(1991)
\item\ M.Y.Choi and J.Choi,\textit{Phys.Rev.B},\textbf{63},212503(2001)
\item\ M.Y.Choi,B.J.Kim and S.Kim,\textit{Phys.Rev.B},\textbf{47},9112(1993)
\item\ H.R.Zeller and I.Giaver,\textit{Phys.Rev.},\textbf{181},789(1969)
\item\ M.Yoneda, M,Niwa and S. Shinohara,\textit{J.Adv.Sci},\textbf{Vol.8,No.3 \& 4},180(1996)\\
(in Japanese)
\item\ H.Kleinert,\textit{GAUGE FIELDS IN CONDENSED MATTER Vol.I}\\
(World Scientific,Singapore 1989)
\item\ H.Kleinert,\textit{Lett.Nuovo Cimento},\textbf{35},405(1982)
\item\ M.Yoneda,M.Niwa and S.Shinohara,\textit{J.Adv.Sci},\textbf{Vol.9 No.3 \& 4},211(1997)\\
(in Japanese)
\item\ M.Yoneda,M.Niwa and S.Shinohara,\textit{J.Adv.Sci},\textbf{Vol.9 No3 \& 4},217(1997)\\
(in Japanese)
\item\ D.Lee and M.P.A.Fisher,\textit{International Journal of Modern Phys.B},\textbf{Vol.5,No.16 \& 17},2675(1991)
\item\ M.Sugahara,\textit{Jpn.J.Appl.Phys.},\textbf{24},674(1985)
\item\ M.Sugahara and N.Yohikawa,\textit{Prog.Theor.Phys.},\textbf{78},957(1987)
\item\ M.Cha,M.P.A.Fisher,S.M.Girvin,M.Wallin and A.P.Young,\textit{Phys.Rev.},\textbf{44},6883(1991)
\item\ S.M.Girvin,M.Wallin,M.C.Cha,M.P.A.Fisher and A.P.Young,\textit{Prog.Theor.Phys.Suppl.}\\
\textbf{No.107},135(1992)
\item\ B.J.van Wees,\textit{Phys.Rev.B},\textbf{44},2264(1991)
\item\ M.P.A.Fisher,\textit{Phys.Rev.Lett.},\textbf{65},923(1990)
\end{enumerate}
\end{document}